\documentclass[11pt]{article}

\usepackage{indentfirst}
\usepackage{amsmath}
\usepackage{amssymb}
\usepackage{amsfonts}
\usepackage[dvipsnames]{xcolor}
\usepackage{multirow}
\usepackage{makecell}
\usepackage{float}

\usepackage{geometry}
\usepackage[backend=bibtex, style=numeric-comp, sorting=none, maxnames=99]{biblatex}
\usepackage{hyperref}
\hypersetup
{
    colorlinks = true,
    linkcolor = red,
    citecolor = blue,
    urlcolor = PineGreen
}

\begin{document}

\begin{titlepage} \setcounter{page}{0}
\begin{center}
    \vspace*{1.0cm}
    {\Large\bf Unbounded Radius of Innermost Stable Circular Orbit in Higher-Dimensional Black Holes}
    \\ \vspace{2.0cm}
    {$\mbox{Hocheol Lee}$}\footnote{\it email: insaying@dongguk.edu}, \quad
    {$\mbox{Bogeun Gwak}$}\footnote{\it email: rasenis@dgu.ac.kr}
    \\ \vspace{0.2cm}
    {\small \it Department of Physics, Dongguk University, Seoul 04620, Republic of Korea}
    \\ \vspace{2.0cm}
\end{center}

\begin{center}
\begin{abstract}
    The innermost stable circular orbit (ISCO) offers a fundamental test of spacetime structure. However, its behavior in higher-dimensional black holes influenced by anisotropic energy-momentum tensors remains insufficiently explored. In this work, we investigate the upper bound of the ISCO in higher-dimensional, static, spherically symmetric, and asymptotically flat black hole spacetimes in the presence of an anisotropic energy-momentum tensor. The energy-momentum tensor is assumed to satisfy the weak energy condition, possess a non-positive trace, and obey constraints on radial and tangential pressures, collectively equivalent to the dominant energy condition with additional constraints. By analyzing the effective potential for timelike geodesics and imposing ISCO conditions, we demonstrate the general absence of an upper bound on the ISCO radius in higher-dimensional spacetimes. For dimensions greater than or equal to eight, an ISCO may not exist, depending on the radial and tangential components of the energy-momentum tensor. If an ISCO exists, its radius remains unbounded. These findings advance our understanding of orbital stability in higher-dimensional gravitational systems and highlight fundamental differences from four-dimensional black hole dynamics.
\end{abstract}
\end{center}
\end{titlepage}

\newpage
\section{Introduction} \label{sec:introduction}
    Black holes have emerged as indispensable theoretical and observational tools for probing the nature of gravity. Originally, they were conceived as idealized solutions to the Einstein field equations~\cite{Schwarzschild:1916uq}. However, they have now been confirmed as genuine astrophysical objects, supported by extensive observational evidence. Observations of supermassive black holes at galactic centers, notably M87$^*$~\cite{Young:1978} and Sagittarius A$^*$~\cite{Schodel:2002py}, offer compelling evidence of the existence of event horizons and relativistic gravitational dynamics. The Event Horizon Telescope’s direct imaging of black hole shadows~\cite{EventHorizonTelescope:2019dse, EventHorizonTelescope:2022wkp} delivers the strongest empirical evidence for the existence of black holes. The image of the black hole shadow allows observational access to spacetime geometry and gravitational effects, confirming the presence of an event horizon, extreme curvature, and strong gravity in its vicinity. Complementary evidence of the existence of black holes arises from stellar-mass compact objects in X-ray binaries, specifically Cygnus X-1~\cite{Bowyer:1965hxq}. The high orbital velocity of the companion star and absence of detectable surface emissions indicate that the compact object in this system possesses a mass that exceeds the theoretical limit for a neutron star~\cite{Orosz:2011np, Gou:2011nq}. Furthermore, the detection of gravitational waves by the LIGO and Virgo collaboration ~\cite{LIGOScientific:2016aoc} revealed the coalescence of binary black holes and constituted the first direct observation of gravitational waves, confirming the century-old prediction of general relativity~\cite{Einstein:1916cc}. The successful detection and analysis of the merger and ringdown phases demonstrate the existence of black holes. These observations offer independent confirmation of the existence of black holes and allow precise measurements of their fundamental properties.

    Among the notable features of black hole spacetimes, the innermost stable circular orbit (ISCO) represents the smallest radius at which a massive particle can maintain a stable circular trajectory. The ISCO defines the inner edge of the standard accretion disk and governs the key physical processes of the black hole accretion system. The ISCO is closely associated with regions where relativistic jets originate, establishing a connection between the inner disk dynamics and jet formation~\cite{Blandford:1977ds}. Measurements of the X-ray binaries reveal features arising from the innermost accretion region in the vicinity of the ISCO, including characteristic shifts in the spectral hardness~\cite{Tananbaum:1972}, quasi-periodic oscillations~\cite{vanderKlis:1985, Middleditch:1986}, and a broadened iron K$\alpha$ line~\cite{Tanaka:1995en}. Furthermore, the radius of the ISCO is essential for inferring the black hole spin and radiative efficiency~\cite{Novikov:1973kta, Page:1974he}. In compact binaries, the ISCO parameters are critical for numerical relativity simulations of the highly curved spacetime region. Additionally, various methods have been developed to determine the ISCO location, including Hamiltonian approaches~\cite{Cook:1994va, Brandt:1997tf, Baumgarte:2000mk}, Pad\`e approximations~\cite{Damour:1997ub}, post-Newtonian approximations~\cite{Clark:1977, Lai:1996sv, Shibata:1997mqo, Favata:2010yd, Blanchet:2025agj}, and variational principles~\cite{Blackburn:1992rr}. However, numerous studies have been conducted on the ISCO around black holes. The ISCO radius for Schwarzschild black holes is three times the event horizon radius. Numerous studies have been conducted for various other spacetime backgrounds~\cite{Suzuki:1997by, Favata:2010ic, Pugliese:2011py, Lee:2017fbq, Schroven:2017jsp, Zhang:2017nhl, Igata:2019hkz, Guo:2020zmf, Schroven:2020ltb, Larranaga:2020ycg, Li:2022sjb, Du:2024ujg, Zhang:2025lhm, Wang:2025bjf}.
    
    Recent studies have shown that a universal upper bound exists for the ISCO in four-dimensional spacetimes~\cite{Cen:2025atm}. However, higher-dimensional black holes remain largely unexplored in this context, and no corresponding universal bounds have yet been identified. Higher-dimensional black holes have significant implications for extensions of general relativity. These implications arise in theories with higher-curvature corrections, unified frameworks incorporating extra dimensions, and braneworld scenarios, where additional spatial dimensions modify gravitational dynamics. These extra dimensions alter the balance between gravitational attraction and centrifugal forces that govern orbital stability in the four dimensions, resulting in different behavior for circular orbits in higher-dimensional spacetimes. Determining whether universal limits on the ISCO persist in higher dimensions is essential for understanding how spacetime geometry and higher dimensions affect orbital stability in strong gravitational fields. Beyond its theoretical significance, the study of orbital stability in higher-dimensional spacetimes offers an observationally accessible probe for testing the predictions of general relativity. This approach further enables the investigation of extra spatial dimensions through astrophysical measurements, allowing the verification of relativistic predictions in extended gravitational theories.

    In this study, we examine the existence and possible upper bound of the ISCO radius in higher-dimensional, static, spherically symmetric, and asymptotically flat spacetimes, using a general and model-independent framework. The analysis considers the physical constraints imposed by the energy-momentum tensor. Additionally, we examine circular orbits in higher-dimensional spacetimes with matter, investigating their existence and stability. We determine the ISCO condition by analyzing equatorial time-like geodesics and conserved quantities in higher-dimensional spacetimes, accounting for anisotropic pressures. Our analysis confirms the existence of a finite upper bound in four dimensions~\cite{Cen:2025atm}. However, in higher dimensions, the ISCO radius may be unbounded or may not exist, depending on the matter content and spacetime dimensions.
    
    The remainder of this paper is organized as follows: Section~\ref{sec:review} reviews ISCO in four-dimensional Schwarzschild black holes. Section~\ref{sec:background} presents the construction of higher-dimensional black hole spacetimes and the associated energy-momentum conditions. Section~\ref{sec:ISCO} investigates the existence and possible upper bounds of the ISCO in higher dimensions. Finally, Section~\ref{sec:conclusion} summarizes the main results and outlines possible directions for future research.

\section{Review of ISCO in Schwarzschild Black Hole Background} \label{sec:review}
    This review examines the existence of an ISCO in the Schwarzschild black hole background. The metric of the Schwarzschild black hole is expressed as
\begin{equation}
    \mathrm{d}s^2 = - h(r) \mathrm{d}t^2 + \frac{\mathrm{d}r^2}{h(r)} + r^2 \mathrm{d}\theta^2 + r^2 \sin^2\theta \mathrm{d}\phi^2,
    \label{review:metric}
\end{equation}
    where
\begin{equation}
    h(r) = 1 - \frac{2 m}{r},
\end{equation}
    with the speed of light set to $c = 1$, and $m$ is mass parameter of the Schwarzschild black hole. We introduce the equatorial motion $( \theta = \pi / 2 )$ of a particle along a timelike geodesic with a normalized four-velocity $\mathbf{u}$, defined as
\begin{equation}
    \mathbf{v} = \left\{ \dot{t}, \dot{r}, 0, \dot{\phi} \right\},
\end{equation}
    where the dot denotes the derivative with respect to the affine parameter. In the equatorial plane, the timelike geodesic $\mathbf{v}^2 = -1$ reduces to
\begin{equation}
    - h \dot{t}^2 + \frac{\dot{r}^2}{h} + r^2 \dot{\phi}^2 = -1.
    \label{review:geodesic_equation}
\end{equation}
    Owing to the static and spherically symmetric metric~\eqref{review:metric}, the two conserved quantities are given by the timelike Killing vector $\zeta^{(t)}_\mu$ and the rotational Killing vector $\zeta^{(\phi)}_\mu$ as follows:
\begin{equation}
    E \equiv - \zeta^{(t)}_\mu v^\mu = h \dot{t}, \qquad L \equiv \zeta^{(\phi)}_\mu v^\mu = r^2 \dot{\phi},
    \label{review:conserved_quantities}
\end{equation}
    where $\boldsymbol{\zeta^{(t)}} = \left\{ 1, 0, 0, 0 \right\}$, $\boldsymbol{\zeta^{(\phi)}} = \left\{ 0, 0, 0, 1 \right\}$, with $E$ and $L$ representing the energy and angular momentum per unit rest mass of a particle, respectively. From the conserved quantities~\eqref{review:conserved_quantities}, the timelike geodesic equation~\eqref{review:geodesic_equation} can be expressed as
\begin{equation}
    \frac{\dot{r}^2}{2} + V_{\mathrm{eff}} = 0,
\end{equation}
    where the effective potential is given by
\begin{equation}
    V_\mathrm{eff}(r) = - \frac{E^2}{2} + \frac{1}{2} \left( 1 + \frac{L^2}{r^2} \right) h(r).
\end{equation}
    The radius of the ISCO, hereafter referred to as $r_\mathrm{ISCO}$, corresponds to a circular orbit that satisfies
\begin{equation}
    V_\mathrm{eff}(r_\mathrm{ISCO}) = 0, \qquad V_\mathrm{eff}'(r_\mathrm{ISCO}) = 0,
    \label{review:ISCO_potential}
\end{equation}
    which determine the conserved quantities as
\begin{equation}
    E^2 = \left. \frac{2 h^2}{2 h - r h'} \right|_{r = r_\mathrm{ISCO}}, \qquad L^2 = \left. \frac{r^3 h'}{2 h - r h'} \right|_{r = r_\mathrm{ISCO}},
\end{equation}
    where the prime denotes the derivative with respect to the radial coordinate $r$. Additionally, the ISCO is defined by the condition that the second derivative of the effective potential vanishes,
\begin{eqnarray}
    0 &=& V_\mathrm{eff}''(r_\mathrm{ISCO})
    \\
    &=& \left. \frac{h''}{2} + \frac{L^2}{r^2} \left( \frac{h''}{2} - \frac{2 h'}{r} + \frac{3 h}{r^2} \right) \right|_{r = r_\mathrm{ISCO}}
    \\
    &=& \left. \frac{\left( r - 6 m \right) m}{r^3 \left( r - 3 m \right)} \right|_{r = r_\mathrm{ISCO}},
\end{eqnarray}
    which results in
\begin{equation}
    r_\mathrm{ISCO} = 6 m.
\end{equation}
    Thus, the ISCO in Schwarzschild spacetime is located at $r_\mathrm{ISCO} = 6 m$, corresponding to three times the event horizon of the Schwarzschild black hole.

\section{System of Higher-Dimensional Black Hole Backgrounds} \label{sec:background}
    To investigate the ISCO in higher-dimensional spacetimes with matter fields, the analysis begins with the $D$-dimensional Einstein-Hilbert action, including matter fields:
\begin{equation}
    S = \int \mathrm{d}^Dx \sqrt{-g} \left( \frac{R}{2 \kappa} + \mathcal{L}_{\mathrm{M}} \right),
\end{equation}
    where $\kappa = 8 \pi G$, $G$ denotes the $D$-dimensional gravitational constant and $\mathcal{L}_\mathrm{M}$ denotes the Lagrangian density of matter. The most general form of the static and spherically symmetric $D$-dimensional metric can be written as
\begin{equation}
    \mathrm{d}s^2 = - e^{2 g(r)} f(r) \mathrm{d}t^2 + \frac{\mathrm{d}r^2}{f(r)} + r^2 \mathrm{d}\Omega_{D-2}^2,
    \label{eq:metric}
\end{equation}
    where $\mathrm{d}\Omega_{D - 2}^2$ is the line element of the unit $(D-2)$-sphere,
\begin{equation}
    \mathrm{d}\Omega_{D - 2}^2 = \mathrm{d}\theta_1^2 + \sin^2\theta_1 \mathrm{d}\theta_2^2 + \cdots + \sin^2\theta_1 \cdots  \sin^2\theta_{D - 3} \mathrm{d}\theta_{D - 2}^2.
\end{equation}
    The asymptotic flatness condition requires
\begin{equation}
    \lim_{r \to \infty} f(r) = 1, \qquad \lim_{r \to \infty} g(r) = 0.
\end{equation}
    We consider a static, diagonal, energy-momentum tensor of an anisotropic fluid, expressed as
\begin{equation}
    \left( T^{\mu}{}_{\nu} \right) = \mathrm{diag} ( - \rho(r), \, p_\mathrm{r}(r), \, \underbrace{p_\mathrm{t}(r), \, p_\mathrm{t}(r), \,\, \cdots\!, \, p_\mathrm{t}(r)}_{D - 2 \textrm{ terms}} ),
\end{equation}
    where $p_\mathrm{r}$ and $p_\mathrm{t}$ are the radial and transverse pressure, respectively. From the Einstein equation, $G^{\mu}{}_{\nu} = \kappa T^{\mu}{}_{\nu}$, the time and radial components are
\begin{eqnarray}
    0 &=& f' - \frac{\left( D - 3 \right) \left( 1 - f \right)}{r} + \frac{2 \kappa r \rho}{D - 2},
    \label{eq:f'}
    \\
    0 &=& g' - \frac{\kappa r \left( \rho + p_\mathrm{r} \right)}{\left( D - 2 \right) f}.
    \label{eq:g'}
\end{eqnarray}
    The second derivatives of Eqs.~\eqref{eq:f'}~and~\eqref{eq:g'} reduce to
\begin{eqnarray}
    f'' &=& - \frac{\left( D - 3 \right) \left( r f' - f + 1 \right)}{r^2} - \frac{2 \kappa \left( r \rho' + \rho \right)}{D - 2},
    \label{eq:f''}
    \\
    g'' &=& - \frac{\kappa \left\{ r \left( \rho + p_\mathrm{r} \right) f' - \left[ r \left( \rho' + p_\mathrm{r}' \right) + \rho + p_\mathrm{r} \right] f \right\}}{\left( D - 2 \right) f^2}.
    \label{eq:g''}
\end{eqnarray}
    The metric function $f(r)$ vanishes at the event horizon $r_\mathrm{h}$ and remains non-negative outside the horizon, satisfying
\begin{equation}
    f(r_\mathrm{h}) = 0, \qquad f'(r_\mathrm{h}) \geq 0.
    \label{eq:f_horizon}
\end{equation}
    The metric function at the event horizon~\eqref{eq:f_horizon} gives
\begin{equation}
    \rho(r_\mathrm{h}) \leq \frac{\left( D - 3 \right) \left( D - 2 \right)}{2 \kappa r_\mathrm{h}^2}, \qquad \rho(r_\mathrm{h}) = - p_\mathrm{r}(r_\mathrm{h}),
    \label{eq:EMT_horizon}
\end{equation}
    where the first inequality is derived from Eq.~\eqref{eq:f'} and the second obtained from Eq.~\eqref{eq:g'}. The metric function $f(r)$ is assumed to take the form
\begin{equation}
    f(r) \equiv 1 - \frac{2 M(r)}{r^{D - 3}},
    \label{eq:f}
\end{equation}
    where $\mu(r)$ characterizes the spacetime curvature generated by the mass distribution outside the horizon. From Eq.~\eqref{eq:f'}, the function $\mu(r)$ can be written as
\begin{equation}
    M(r) = \frac{r_\mathrm{h}^{D - 3}}{2} + \frac{\kappa}{D - 2} \int_{r_\mathrm{h}}^{r} s^{D - 2} \rho(s) \mathrm{d}s.
\end{equation}
    The finiteness of $M(r)$ imposes the following condition on energy density $\rho(r)$:
\begin{equation}
    \lim_{r \to \infty} r^{D - 1} \rho(r) = 0.
    \label{eq:asymptotic_condition}
\end{equation}
    From the energy-momentum conservation law $\nabla_\mu T^{\mu}{}_{r} = 0$, and by substituting Eqs.~\eqref{eq:f'} and~\eqref{eq:g'}, the radial derivative of the pressure is expressed as
\begin{equation}
    p_\mathrm{r}' = \frac{2 \mathbf{T} + \left( D - 1 \right) \rho - \left( D + 1 \right) p_\mathrm{r}}{2 r} - \left( \frac{D - 3}{2 r} + \frac{\kappa r p_\mathrm{r}}{D - 2} \right) \frac{\rho + p_\mathrm{r}}{f},
    \label{eq:conservation}
\end{equation}
    where $\mathbf{T}$ denotes the trace of the energy-momentum tensor $T^{\mu}{}_{\mu}$. The energy-momentum conditions are imposed as follows: \textbf{(\romannumeral 1)} The energy-momentum tensor is required to satisfy the weak energy condition:
\begin{equation}
    \rho \geq 0, \qquad \rho + p_\mathrm{r} \geq 0, \qquad \rho + p_\mathrm{t} \geq 0, 
\end{equation}    
    \textbf{(\romannumeral 2)} the trace of the energy-momentum tensor is assumed to be non-positive,
\begin{equation}
    \mathbf{T} \leq 0,
\end{equation}
    which implies a non-negative scalar curvature, and the non-positive trace condition can be expressed as
\begin{equation}
    \rho \geq p_\mathrm{r} + \left( D - 2 \right) p_\mathrm{t},
\end{equation}
    \textbf{(\romannumeral 3)} the radial and tangential pressures are taken to be
\begin{equation}
    p_\mathrm{t} \geq 0, \qquad p_\mathrm{t} \geq \left| p_\mathrm{r} \right|,
\end{equation}
    which ensures that the tangential pressure does not act as tension, suppresses excessive radial pressure, and prevents violations of the energy-momentum conditions. Combining the conditions \textbf{(\romannumeral 1)}, \textbf{(\romannumeral 2)}, and \textbf{(\romannumeral 3)} results in a dominant energy condition accompanied by the following additional constraints:
\begin{equation}
   \rho \geq \left| p_\mathrm{r} \right|, \qquad \rho \geq \left| p_\mathrm{t} \right|, \qquad \rho \geq p_\mathrm{r} + \left( D - 2 \right) p_\mathrm{t} \geq 0, \qquad p_\mathrm{t} \geq \left| p_\mathrm{r} \right| \geq 0.
   \label{eq:energy_conditions}
\end{equation}

\section{ISCO in Higher-Dimensional Black Hole Backgrounds} \label{sec:ISCO}
    We consider the $D$-dimensional equatorial motion of a particle along a timelike geodesic with a $D$-dimensional normalized four-velocity $\mathbf{u}$, defined as
\begin{equation}
    \mathbf{u} = \left\{ \dot{t}, \dot{r}, \dot{\theta}_1, \cdots, \dot{\theta}_{D-2} \right\}.
\end{equation}
    The $D$-dimensional equatorial plane is specified by fixing all the polar angles at $\pi / 2$ as follows:
\begin{equation}
    \theta_1 = \theta_2 = \cdots = \theta_{D-3} = \frac{\pi}{2}, \qquad \theta_{D-2} \equiv \phi,
\end{equation}
    where $\phi$ denotes redefined azimuthal coordinates. The energy $E$ and angular momentum $L$ are given by the $D$-dimensional timelike Killing vector $\xi^{(t)}_\mu$ and the rotational Killing vector $\xi^{(\phi)}_\mu$,
\begin{equation}
    E \equiv - \xi^{(t)}_\mu u^\mu = e^{2 g} f \dot{t}, \qquad L \equiv \xi^{(\phi)}_\mu u^\mu = r^2 \dot{\phi},
    \label{eq:energy_angular_momentum}
\end{equation}
    where $\boldsymbol{\xi^{(t)}} = \left\{ 1, 0, \cdots, 0 \right\}$, $\boldsymbol{\xi^{(\phi)}} = \left\{ 0, \cdots, 0, 1 \right\}$. In the equatorial plane, the timelike geodesic $\mathbf{u}^2 = -1$ with two conserved quantities~\eqref{eq:energy_angular_momentum} reduces to
\begin{equation}
    \frac{\dot{r}^2}{2} + \frac{f}{2} \left( - \frac{E^2 e^{- 2 g}}{f} + \frac{L^2}{r^2} + 1 \right) = 0.
    \label{eq:geodesic_equation}
\end{equation}
    From the timelike geodesic~\eqref{eq:geodesic_equation}, the effective potential is defined as
\begin{equation}
    V_\mathrm{eff}(r) \equiv - \frac{E^2}{2} e^{-2 g(r)} + \frac{1}{2} \left( 1 + \frac{L^2}{r^2} \right) f(r).
\end{equation}
    The energy $E$ and angular momentum $L$ are determined by the ISCO condition~\eqref{review:ISCO_potential},
\begin{equation}
    E^2 = \left. \frac{2 e^{2 g} f^2}{2 f - r \left( f' + 2 f g' \right)} \right|_{r = r_\mathrm{ISCO}}, \qquad L^2 = \left. \frac{r^3 \left( f' + 2 f g' \right)}{2 f - r \left( f' + 2 f g' \right)} \right|_{r = r_\mathrm{ISCO}}.
    \label{eq:E&L}
\end{equation}
    Applying Eqs.~\eqref{eq:f'},~\eqref{eq:g'},~\eqref{eq:f''},~\eqref{eq:g''},~\eqref{eq:conservation},~and~\eqref{eq:E&L}, the condition $V''(r_\mathrm{ISCO})  = 0$ reduces to
\begin{eqnarray}
    0 &=& \frac{2 \kappa^2 r^4 p_\mathrm{r} \left( \rho - 3 p_\mathrm{r} \right)}{\left( D - 2 \right)^2} + \frac{\kappa r^2 \left\{ 2 f \mathbf{T} + \left[ \left( D - 1 \right) f + D - 3 \right] \rho + \left[ \left( 5 D - 11 \right) f - 7 \left( D - 3 \right) \right] p_\mathrm{r} \right\}}{D - 2} \nonumber
    \\
    && \quad \left. - \left( D - 3 \right) \left( D - 1 \right) f^2 + \left( 3 D - 7 \right) \left( D - 3 \right) f - 2 \left( D - 3 \right)^2 \right|_{r = r_\mathrm{ISCO}}.
    \label{eq:ISCO}
\end{eqnarray}
    The condition~\eqref{eq:ISCO} can be reformulated to $\mathcal{C}(r)$ for an arbitrary radius $r$, defined as
\begin{eqnarray}
    \mathcal{C}(r) \!\!\! &\equiv& \!\!\! \frac{2 \kappa^2 r^4 p_\mathrm{r} \left( \rho - 3 p_\mathrm{r} \right)}{\left( D - 2 \right)^2} + \frac{\kappa r^2 \left\{ 2 f \mathbf{T} + \left[ \left( D - 1 \right) f + D - 3 \right] \rho + \left[ \left( 5 D - 11 \right) f - 7 \left( D - 3 \right) \right] p_\mathrm{r} \right\}}{D - 2} \nonumber
    \\
    && \quad - \left( D - 3 \right) \left( D - 1 \right) f^2 + \left( 3 D - 7 \right) \left( D - 3 \right) f - 2 \left( D - 3 \right)^2,
    \label{eq:W}
\end{eqnarray}
    and the function $\mathcal{C}(r)$ can be decomposed into
\begin{equation}
    \mathcal{C}(r) \equiv \mathcal{A}(r) + \mathcal{B}(r),
\end{equation}
    with
\begin{eqnarray}
    \mathcal{A}(r) \!\!\!\! &=& \!\!\!\! \frac{2 \kappa^2 r^4 p_\mathrm{r} \left( \rho - 3 p_\mathrm{r} \right)}{\left( D - 2 \right)^2} + \frac{\kappa r^2 \left\{ 2 f \mathbf{T} + \left[ \left( D - 1 \right) f + D - 3 \right] \rho + \left[ \left( 5 D - 11 \right) f - 7 \left( D - 3 \right) \right] p_\mathrm{r} \right\}}{D - 2} ,
    \\
    \mathcal{B}(r) \!\!\!\! &=& \!\!\!\! - \left( D - 3 \right) \left( D - 1 \right) f^2 + \left( 3 D - 7 \right) \left( D - 3 \right) f - 2 \left( D - 3 \right)^2.
\end{eqnarray}
    Considering the condition~\eqref{eq:EMT_horizon}~and~energy-momentum conditions~\eqref{eq:energy_conditions}, the function $\mathcal{C}(r)$ at the horizon satisfies
\begin{equation}
    \mathcal{C}(r_\mathrm{h}) = - \frac{2 \left[ 2 \kappa r^2 \rho - \left( D - 3 \right) \left( D - 2 \right) \right]^2}{\left( D - 2 \right)^2} \leq 0,
\end{equation}
    and in the asymptotic region, with conditions~\eqref{eq:asymptotic_condition}~and~\eqref{eq:energy_conditions}, $\mathcal{C}(r)$ satisfies
\begin{equation}
    \mathcal{C}(r \to \infty) \sim \frac{2 \kappa r \left[ \left( D - 3 \right) \rho - \left( D - 6 \right) p_\mathrm{r} + \left( D - 2 \right) p_\mathrm{t} \right]}{D - 2} \geq 0.
\end{equation}
    As $\mathcal{C}(r)$ is continuous, $r_\mathrm{ISCO}$ exists in the range $r_\mathrm{h} < r_\mathrm{ISCO} < \infty$ for which
\begin{equation}
    \mathcal{C}(r_\mathrm{ISCO}) = 0.
\end{equation}
    The non-negativity of energy and angular momentum of the particle~\eqref{eq:E&L} imposes the condition
\begin{equation}
    \frac{2 \kappa r^2 p_\mathrm{r} + \left( D - 3 \right) \left( D - 2 \right)}{\left( D - 2 \right) \left( D - 1 \right)} < f(r) \leq 1 + \frac{2 \kappa r^2 p_\mathrm{r}}{\left( D - 3 \right) \left( D - 2 \right)},
    \label{eq:condition_f}
\end{equation}
    and within the domain $r_\mathrm{h} < r < \infty$, where $0 < f(r) < 1$, the following inequality holds:
\begin{equation}
    0 < 1 + \frac{2 \kappa r^2 p_\mathrm{r}}{\left( D - 3 \right) \left( D - 2 \right)}.
\end{equation}
    To simplify the subsequent analysis, we introduce two quantities
\begin{equation}
    \delta(r) \equiv 1 + \frac{2 \kappa r^2 p_\mathrm{r}}{\left( D - 3 \right) \left( D - 2 \right)} > 0, \qquad \sigma(r) \equiv p_\mathrm{r} + p_\mathrm{t} \geq 0.
\end{equation}

\subsection{Positive Radial Pressure} \label{sec:positive_pressure}
    The analysis can be separated according to the sign of $p_\mathrm{r}$. Regarding a positive radial pressure $( p_\mathrm{r} > 0 )$, the relation $\delta > 1$ holds, and from the energy-momentum conditions~\eqref{eq:energy_conditions} and condition~\eqref{eq:condition_f}, we obtain
\begin{eqnarray}
    \mathcal{A}(r) &\geq& \frac{4 \kappa^2 r^4 p_\mathrm{r} \left\{ \left( D - 4 \right) \left( D - 1 \right) \rho + \left[ \left( D - 3 \right)^2 + 2 \right] p_\mathrm{r} + \left( D - 2 \right) \left( D - 1 \right) p_\mathrm{t} \right\}}{\left( D - 3 \right) \left( D - 2 \right)^2 \left( D - 1 \right)} \nonumber
    \\
    && \, + \frac{2 \kappa r^2 \left\{ \left( D - 4 \right) \left( D - 1 \right) \rho - \left[ \left( D - 1 \right)^2 - 6 \right] p_\mathrm{r} + \left( D - 2 \right) \left( D - 1 \right) p_\mathrm{t} \right\}}{\left( D - 2 \right) \left( D - 1 \right)}
    \\
    &=& \frac{2 \left( D - 4 \right) \kappa r^2 \delta }{D - 2} + \frac{4 \kappa^2 r^4 p_\mathrm{r} \left\{ \left[ \left( D - 3 \right)^2 + 2 \right] \sigma + 3 \left( D - 3 \right) p_\mathrm{t} \right\}}{\left( D - 3 \right) \left( D - 2 \right)^2 \left( D - 1 \right)} \nonumber
    \\
    && \, + \frac{2 \kappa r^2 \left\{ - \left[ \left( D - 1 \right)^2 - 6 \right] p_\mathrm{r} + \left( D - 2 \right) \left( D - 1 \right) p_\mathrm{t} \right\}}{\left( D - 2 \right) \left( D - 1 \right)}
    \\
    &\geq& \frac{2 \kappa r^2 \left[ - \left( D - 7 \right) p_\mathrm{r} + \left( D - 2 \right) \left( D - 1 \right) \left( p_\mathrm{t} - p_\mathrm{r} \right) \right]}{\left( D - 2 \right) \left( D - 1 \right)}
    \\
    &\geq& - \frac{2 \left( D - 7 \right) \kappa r^2 p_\mathrm{r}}{\left( D - 2 \right) \left( D - 1 \right)}.
\end{eqnarray}
    For dimensions $D \leq 7$, the function $\mathcal{A}(r)$ remains non-negative. The continuity of $\mathcal{C}(r)$ from negative to positive values, combined with $\mathcal{A}(r) \geq 0$, implies the condition $\mathcal{C}(r_\mathrm{ISCO}) = 0$ requires $\mathcal{B}(r_\mathrm{ISCO}) \leq 0$, yielding
\begin{eqnarray}
    \frac{2 M(r_\mathrm{ISCO})}{r_\mathrm{ISCO}^{D - 3}} \geq - \frac{D - 5}{D - 1}.
    \label{eq:ISCO_condition}
\end{eqnarray}
    In the four-dimensional case $( D = 4 )$, the inequality~\eqref{eq:ISCO_condition} gives the upper bound on $r_\mathrm{ISCO}$,
\begin{eqnarray}
    r_\mathrm{ISCO} \leq 6 M,
\end{eqnarray}
    which is consistent with the result of~\cite{Cen:2025atm}. For $5 \leq D \leq 7$, the inequality~\eqref{eq:ISCO_condition} reduces to
\begin{eqnarray}
    r_\mathrm{ISCO} > 0,
    \label{eq:unbounded}
\end{eqnarray}
    indicating the absence of the upper bound on $r_\mathrm{ISCO}$. In contrast, when the number of dimension exceeds seven $( D \geq 8 )$, the sign of $\mathcal{A}(r)$ can be either positive or negative. If $\mathcal{A}(r) \geq 0$, the identical form of the expressions allows the argument employed in condition~\eqref{eq:unbounded} to establish that $r_\mathrm{ISCO}$ is unbounded, whereas for $\mathcal{A}(r) < 0$, the condition $\mathcal{B}(r) > 0$ leads to
\begin{eqnarray}
    r_\mathrm{ISCO}^{D - 3} < - \frac{2 \left( D - 1 \right) M(r_\mathrm{ISCO})}{D - 5} < 0,
\end{eqnarray}
    demonstrating the absence of $r_\mathrm{ISCO}$.

\subsection{Non-Positive Radial Pressure} \label{sec:non-positive_pressure}
    For the non-positive radial pressure $(p_\mathrm{r} \leq 0)$, where $0 < \delta \leq 1$, we find that
\begin{eqnarray}
    \mathcal{A}(r) \!\!\! &\geq& \!\!\! \frac{4 \kappa^2 r^4 p_\mathrm{r} \left[ \left( D - 4 \right) \rho + D p_\mathrm{r} + \left( D - 2 \right) p_\mathrm{t} \right]}{\left( D - 3 \right) \left( D - 2 \right)^2} + \frac{2 \kappa r^2 \left[ \left( D - 4 \right) \rho - \left( D - 6 \right) p_\mathrm{r} + \left( D - 2 \right) p_\mathrm{t} \right]}{D - 2}
    \\
    &=& \!\!\! \frac{2 \kappa r^2 \delta}{D - 2} \left[ \left( D - 4 \right) \rho + \left( D - 2 \right) \sigma \right] + 2 \left( D - 3 \right) \left( D - 3 - \delta \right) \left( 1 - \delta \right)
    \\
    &\geq& 0,
\end{eqnarray}
    satisfying the dimensions $D \geq 4$. By applying the argument presented for the positive radial-pressure case (Sec. ~\ref{sec:positive_pressure}), the condition $\mathcal{A}(r) \geq 0$ results in an expression equivalent to the inequality~\eqref{eq:ISCO_condition}. Consequently, for $D \geq 5$, the ISCO radius is unbounded from the above, which is identical to the result obtained for a positive radial pressure~\eqref{eq:unbounded}.

\section{Conclusion} \label{sec:conclusion}
    In this study, we investigated the upper bound of the innermost stable circular orbit in higher-dimensional, static, spherically symmetric, and asymptotically flat black hole backgrounds. The anisotropic energy-momentum tensor was considered under the weak energy condition, non-positive trace of the energy-momentum tensor, and constraints on the radial and tangential pressures. The combined requirements are equivalent to the dominant energy conditions with additional constraints.

    For a positive radial pressure, an upper bound on the ISCO radius is realized exclusively in four-dimensional spacetimes. However, in dimensions $5 \leq D \leq 7$, the ISCO radius remains unbounded. For $D \geq 8$, the matter contribution can result in either an unbounded or non-existent ISCO radius. For non-positive radial pressure, except for the four-dimensional case, the ISCO radius is unbounded for all dimensions $D \geq 5$. Table~\ref{tab:r_ISCO} summarizes the overall results.
\begin{table}[H]
    \centering \renewcommand{\arraystretch}{1.5}
    \begin{tabular}{|c|c|c|c|} \hline
        \textbf{Radial Pressure} & \textbf{Dimensions} & $\mathbf{\mathcal{A}(r)}$ & $\mathbf{r}_\mathrm{\textbf{ISCO}}$
        \\ \hline
        \multirow{4}{*}{$p_\mathrm{r} > 0$} & $D = 4$ & $\geq 0$ & $r_\mathrm{ISCO} \leq 6 M$\cite{Cen:2025atm} 
        \\ \cline{2-4}
        & $5 \leq D \leq 7$ & $\geq 0$ & Unbounded
        \\ \cline{2-4}
        & \multirow{2}{*}{$D \geq 8$} & $\geq 0$ & Unbounded 
        \\ \cline{3-4}
        & & $<0$ & Non-existent
        \\ \hline
        \multirow{2}{*}{$p_\mathrm{r} \leq 0$} & $D = 4$ & $\geq 0$ & $r_\mathrm{ISCO} \leq 6 M$\cite{Cen:2025atm}
        \\ \cline{2-4}
        & $D \geq 5$ & $\geq 0$ & Unbounded
        \\ \hline
    \end{tabular}
    \caption{Upper bound and existence of the radius of the innermost stable circular orbit in four and higher dimensions.}
    \label{tab:r_ISCO}
\end{table}
    Therefore, our analysis clarified the dimensional dependence of the ISCO under the dominant energy condition and additional constraints on the pressures. In contrast to the vacuum case, we demonstrated that the presence of an energy-momentum tensor satisfying the dominant energy condition, including additional constraints, allowed the existence of an innermost stable circular orbit in higher-dimensional black hole spacetimes. The results further reveal the absence of an upper bound on the ISCO radius in higher dimensions. Future research may extend the present framework to rotating spacetimes, asymptotically dS\slash AdS backgrounds, or modified gravity to investigate whether the corresponding dimensional dependencies emerge.

\section*{Acknowledgments}
    This research was supported by Basic Science Research Program through the National Research Foundation of Korea (NRF) funded by the Ministry of Education (NRF-2022R1I1A2063176) and the Dongguk University Research Fund of 2025.

\printbibliography

\end{document}